\documentstyle[psfig]{mn}
\input{psfig}
\title{The RMS Peculiar Velocity of Clusters}
\author[Richard Watkins]
{Richard Watkins\\
Department of Physics and Astronomy,
Dartmouth College, Hanover NH 03755, USA\\
rwatkins@dartmouth.edu}

\begin{document}
\maketitle

\begin{abstract}

We study the rms peculiar velocity of clusters as a convenient
statistic to put constraints on cosmological models.  This
statistic is easy to compute theoretically given a model for the
power spectrum; we show that with some assumptions it can be
directly related to the quantity $\Omega_0^{0.6}\sigma_8$.  We
develop a likelihood method for estimating the rms peculiar
velocity of clusters from line-of-sight velocity measurements
and their associated errors.  We apply our method to two samples
of cluster peculiar velocities; a new sample known as the SCI
sample and a subsample of the Mark III catalog.  Although these
two samples initially give results which are inconsistent, we
show that they can be put into good agreement by the removal,
particularly from the Mark III sample, of a set of clusters with
multiple mass concentrations, cluster membership ambiguities, or
excessive obscuration, properties which may have introduced
unaccounted for errors into measurements of their peculiar velocities.
Once these clusters are removed from the samples, they both
favor a relatively low value for the 1-D rms peculiar velocity
of clusters $\sigma_v= 265^{+106}_{-75}{\rm\ km\ s}^{-1}$ (at
$90\%$ confidence), leading to the constraint
$\Omega_0^{0.6}\sigma_8= 0.44^{+0.19}_{-0.13}$, consistent with
cluster abundance studies but inconsistent
with $\Omega_0=1$ CDM normalized to COBE at the
$99.7\%$ confidence level. 

\end{abstract}


\section{Introduction}

Several recent papers have shown that the peculiar velocities of
clusters of galaxies can be used to put strong constraints on
cosmological parameters (Croft and Efstathiou 1994; Bahcall,
Gramann, and Cen 1994; Bahcall and Oh 1996; Moscardini et
al. 1996; Borgani et al. 1997; Shi 1997).  Much of their utility in this
regard derives from the fact that the motions of clusters probe
the peculiar velocity field predominantly on scales where it is
still linear, making the comparison to theoretical models relatively
straightforward.  In addition, peculiar
velocities of clusters can in principle be measured to much higher accuracy
than those of galaxies, since many galaxies
can be measured per cluster.  Thus the motions of clusters
should provide a valuable tool in the study of large scale
structure.  

However, results obtained thus far using different catalogs of
cluster peculiar velocities have been contradictory (see, e.g.,
Moscardini et al. 1996).  In order to have confidence in the use
of cluster peculiar velocities as a probe of large scale
structure it is important to understand why this is so.  In this
{\it Letter} we study this issue, focusing on the estimation of
the rms peculiar velocity of clusters.  This statistic gives us
a convenient way to compare different samples of cluster
peculiar velocities and can be compared directly to predictions
of cosmological models obtained from either $N$-body simulations
or linear theory.  Indeed, as we show below, the rms peculiar
velocity of clusters can be directly related to the combination
of parameters $\Omega_0^{0.6}\sigma_8$, where $\sigma_8$ is the
rms mass fluctuation on a scale of $8\ h^{-1}$Mpc ($H_0 = 100\
h\ {\rm\ km\ s^{-1}\ Mpc^{-1}}$).

We use a likelihood analysis to estimate the rms peculiar
velocity of clusters from line-of-sight velocities and their
estimated errors.  This analysis allows us to assess the
consistency of two samples in a model independent way.  We apply
our method to two well-studied catalogs of cluster
peculiar velocities; a new sample based on measurements of 782
spiral galaxies in the fields of 24 clusters described in
Giovanelli et al. (1997) (hereafter the SCI sample) and a subset
of the cluster peculiar velocities contained in the Mark III
catalog (Willick et al. 1995, 1996, 1997a) based on measurements
of 346 galaxies in the fields of 31 clusters described in Mould
et al. (1991), Han (1992), and Mould et al. (1993) (hereafter
the HM sample).  The HM sample was used to set the zero-point of
the Tully-Fisher relation for the Mark III catalog.  
These samples cover the sky uniformly, have comparable
depths, and are based on Tully-Fisher distance determinations.
As we shall show below, these two samples yield results which
are inconsistent at a high level.

We argue that the inconsistency between the two samples is due
to the inclusion of clusters, mostly in the HM sample, whose
associated galaxy samples are not representative of a
single collapsed object or where measurement difficulties make
error estimates unreliable.  
We show that when clusters suspected of being in this class 
are removed, the SCI and HM samples show excellent agreement,
yielding a value of the rms peculiar velocity of clusters $\sim
250{\rm\ km\ s}^{-1}$.  This favors values of
$\Omega_0^{0.6}\sigma_8\sim 0.5$, in agreement with cluster
abundance studies.  Our results emphasize that the careful
selection of the galaxy sample used in the analysis of a cluster
is crucial for the accurate determination of its peculiar
velocity.

\section{Analysis}

Linear theory of the growth of perturbations suggests that the
1-D peculiar velocity dispersion $\sigma_v^2=\langle v^2_{\rm
1D}\rangle$ of clusters is given by an expression of the form
\begin{equation}
\sigma_v^2 = {1\over 3}{H_o^2\Omega_0^{1.2}\over
2\pi^2}\int_0^\infty\ P(k) W^2(k)\ dk,
\end{equation}
where $P(k)$ is the power spectrum and $W^2(k)$ is a window
function that filters out power on scales smaller than a cluster
size.  Croft and Efstathiou (1994), Bahcall et al. (1994),
and Colberg et al. (1997) have shown that
Eqn. 1 gives good agreement with N-body simulations when a
Gaussian window function is used, $W^2(k)= \exp (-k^2R^2)$,
although they disagree on the best value for the parameter $R$.
We choose to adopt the 
value $R=1.5\ h^{-1}$Mpc used by Colberg et al. (1997), which
works well for models normalized to cluster abundances.
Choosing the larger value of $3.9\ h^{-1}$Mpc used by Borgani et
al. 1997 only changes our results by $\sim 15\%$.  

Assuming a scale-invariant spectrum of initial mass
fluctuations, we follow Efstathiou, Bond, and White (1992)
and parametrize the power spectrum as
\begin{equation}
P(k)= {Bk\over \{1+[ak + (bk)^{3/2} + (ck)^2]^{\nu}\}^{2/\nu}},
\end{equation}
where $a=(6.4/\Gamma)\ h^{-1}$Mpc, $b=(3.0/\Gamma)\ h^{-1}$Mpc,
$c=(1.7/\Gamma )\ h^{-1}$Mpc, and $\nu=1.13$.  Here the
parameter $\Gamma$ ($=\Omega_0h$ in models with $\Omega_0 +
\Lambda =1$) determines the ``shape'' of the power spectrum.
Although motivated by CDM models, this parametrization is able
to fit a a wide range of theoretical models, including those
that include mixed dark matter.

The normalization $B$ can be expressed in terms of the rms
linear mass fluctuation on scales of $8\ h^{-1}$Mpc, $\sigma_8$,
which is defined by
\begin{equation}
\sigma_8^2 \equiv {1\over
(2\pi)^3}\int_0^\infty\ P(k) W_8^2(k)\ d^3k,
\end{equation}
where $W_8(k) ={3j_1(kR_8)/ kR_8}$ with $R_8 = 8\
h^{-1}$Mpc and $j_1$ is a spherical Bessell function.  
Following Colberg et al. 1997, we can express $\sigma_v$ as 
\begin{equation}
\sigma_v = {100{\rm \ km\ s^{-1}}\over \sqrt{3}}\
\Omega_0^{0.6}\sigma_8 \sqrt{f(\Gamma)},
\end{equation}
where the function $f(\Gamma)$ is defined as
\begin{equation}
f(\Gamma)\equiv 4\pi h^2{\int P(k)W^2(k)\ dk\over \int
P(k)W_8^2(k)\ d^3k},
\end{equation}
and is fit well by the equation
$f(\Gamma)=12.5\Gamma^{-1.08}+49.4$.  Thus we can place an upper
bound $\Omega_0^{0.6}\sigma_8 \le \sigma_v(405{\rm\ km\ s}^{-1})^{-1}$.
Alternatively, we note that $\sqrt{f}$ varies only between
$11.0$ and $8.7$ over the range of values $\Gamma= 0.2-0.5$.  If
we make the reasonable assumption that $\Gamma$ lies in this
range (see, e.g., Peacock and Dodds 1994), then to a good
approximation $\sqrt{f}=10$ and $\Omega_0^{0.6}\sigma_8
= \sigma_v(580{\rm\ km\ s}^{-1})^{-1}$.  Colberg et al. (1997) take
a different approach, noting that if one adopts the value of
$\sigma_8$ obtained from cluster abundance studies 
then the quantity $\Omega_0^{0.6}\sigma_8$ varies only very weakly
with $\Omega_0$, so that $\sigma_v$ is nearly a function of
$\Gamma$ alone.  Here we choose to view the measurement of
$\sigma_v$ as an opportunity to independently verify the results
of cluster abundance measurements.

A problem that arises in calculating $\sigma_v$ from 
cluster data is that even if actual peculiar velocities
are gaussian distributed, measured velocities are not.
This is due to the fact that each velocity measurement has its
own distinct error.  For distant objects, these errors can often
be comparable to the velocities themselves and must be properly
accounted for.  Previous studies
have relied on Monte Carlo methods to generate ensembles of
simulated data sets convolved with measurement errors, which
are then compared to the actual data.  Here, we introduce a
likelihood analysis by which the rms peculiar velocity of
clusters can be estimated directly from measured 
line-of-sight velocities and their errors.

We assume that each measured line-of-sight velocity $v_i$ is
drawn from a gaussian distribution with variance $\sigma_v^2 +
\sigma_i^2$, where $\sigma_i$ is the measurement error in the
measurement of $v_i$.  With this assumption, the likelihood of
measuring a set of $v_i$ given a particular value of $\sigma_v$
and measurement errors $\sigma_i$ is given by
\begin{equation}
 L(\sigma_v) =\prod_i {1\over (2\pi)^{1/2}(\sigma_v^2 + \sigma_i^2)^{1/2}}
\exp\left({-v_i^2\over 2(\sigma_v^2 + \sigma_i^2)}\right).
\end{equation}

Bayes' Theorem tells us that the posterior probability
$P(\sigma_v|v_i,\sigma_i)= L(\sigma_v)P(\sigma_v)$, where
$P(\sigma_v)$ is the prior and contains any information that we
might know about $\sigma_v$ beforehand.  In this paper we shall
make the simplest assumption that $P(\sigma_v)=constant$, so
that $P(\sigma_v|v_i,\sigma_i)\propto L(\sigma_v)$.  
For a given set of velocities and errors $(v_i,\sigma_i)$, Eqn. 6
can be used to calculate the maximum likelihood value
of $\sigma_{v}$ as well as confidence intervals.
We shall choose confidence intervals for $\sigma_v$ using the
criterion that $L(\sigma_v)>C$, where $C$ is determined by the
desired confidence level.  

Models of large scale structure can be accepted or rejected by
comparing their predicted value of the peculiar velocity
dispersion, $\sigma_{\rm pred}$ to the likelihood distribution
$L(\sigma_v)$.  In particular, a confidence level for the
rejection of a model can be assessed by calculating the
relative area under the likelihood function satisfying
$L(\sigma_v) \le L(\sigma_{\rm pred})$.
\section{Results}

In Fig. 1 we show the likelihood function $L(\sigma_v)$ for the
SCI and HM cluster samples, both normalized so that they have
unit area.  The relatively low value of $\sigma_v$ favored by
the SCI sample is consistent with the value of $\sigma_v \sim
250{\rm\ km\ s}^{-1}$ predicted by models normalized to reproduce the
correct abundance of clusters (Colberg et al. 1997), and is
inconsistent at the $99.8\%$ confidence level with the value of
$\sigma_v=489{\rm\ km\ s}^{-1}$ predicted by an $\Omega_0=1$ CDM model
normalized to microwave background fluctuations (Bahcall et
al. 1994).  (Note that these papers quote the 3-D rms velocity,
which must be divided by $\sqrt{3}$ for comparison to $\sigma_v$,
the 1-D rms velocity.)  The larger value favored by the HM sample
is quite consistent with $\Omega_0=1$ CDM normalized to COBE
fluctuations.

\begin{figure}
\psfig{figure=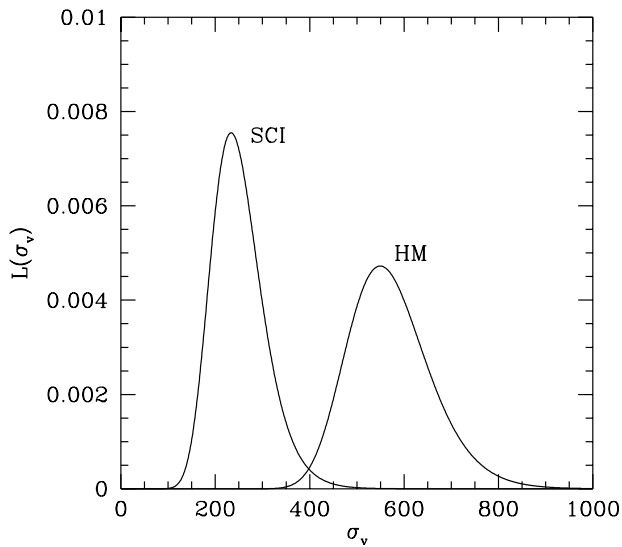,height=3.6in,angle=0}
 \caption{The normalized likelihood functions for $\sigma_v$ obtained from
 the SCI and HM samples.}
\end{figure}

We can quantify the disagreement between the
two samples using their normalized likelihood functions.  The
probability of the two samples having values of $\sigma_v$ which
differ by $\Delta$ is given by $P(\Delta)\propto
\int_0^{\infty} L_{\rm HM}(\Delta+\sigma_v)L_{\rm
SCI}(\sigma_v)\ d\sigma_v$.  A measure of the
probability that the two data sets are drawn from distributions
with the same $\sigma_v$ is given by the relative area under the
curve satisfying $P(\Delta) < P(\Delta=0)$.  Using this measure
we find that the SCI and HM samples disagree on the value of
$\sigma_v$ at the $99.8\%$ confidence level.  

Insight into the nature of the discrepancy between the two
samples can be gained by examining the 12 objects that they have
in common.  In Fig. 2, we show the velocity and error estimates
for these 12 clusters as given in the SCI ($x$-axis) and HM
($y$-axis) catalogs.  The measurements are in good agreement
except for the Cancer cluster, corresponding to the point in the
lower middle of the plot.  This cluster has been studied by
Bothun et al. (1983), who concluded that it consisted of several
discrete groups spread in velocity over $2800{\rm\ km\ s}^{-1}$.
Furthermore, their results suggest that the groups are not
gravitationally bound, but rather are separating with the Hubble
flow.  Giovanelli et al. (1997) selected all of their Cancer
cluster galaxies from the largest mass group, whereas the Cancer
cluster galaxies in the HM sample are drawn from several of the
groups.  The good agreement of the reported velocities for the
less complex clusters suggests that the inconsistency regarding
the Cancer cluster is due to the difference in membership
criteria.  Indeed, it seems likely that the velocity from 
the HM sample does not reflect that of a single gravitationally
bound object.

\begin{figure}
\psfig{figure=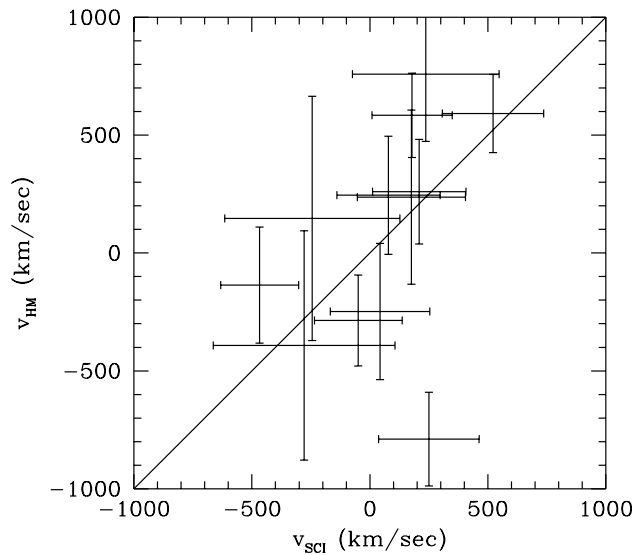,height=3.6in,angle=0}
 \caption{Peculiar velocities and $1\sigma$errors of the 12 objects
 common to the SCI and HM samples.}
\end{figure}

In order to explore the possibility that the discrepancy between
the SCI and the HM samples is due to effects that are not
properly accounted for in the error estimation, we construct
subsamples SCI$_*$ and HM$_*$ of the cluster samples from which
we remove clusters that we consider suspect.  These mostly
consist of clusters which the observers noted as having
ambiguities in their associated galaxy samples or particularly
poor data.  We also remove clusters for which other observers
have noted evidence for multiple mass concentrations or overlap
with other clusters.  From the SCI sample, we excise six paired
clusters (A2197/A2199, Pavo/Pavo II, and A2634/A2666) which have
possible cluster membership ambiguities.  A2634 and Pavo II are
also in the HM sample, and we remove them from this sample as
well.  The following additional clusters are removed from the HM
sample: Cancer (see above), Pisces (Sakai et al. 1994), MKW1S,
and A779 (Mould et al. 1993), due to their having multiple mass
concentrations; 3C 296, due to its distribution of galaxies on
the sky not being well defined (Mould et al. 1993); 1559+19, due
to it being sparsely sampled and composed of rather
inhomogenious data (Mould et al. 1993); OCA 3627, due to
difficulties arising from its low galactic latitude (Mould et
al. 1991); N5419, due to poor sampling and possible substructure
(Mould et al. 1991); and OCA 3560 (Mould et al. 1991), ESO 508,
and CEN 45 (Aaronson et al. 1989), due to possible contamination
by foreground or background galaxies.  This leaves us with
reduced samples of 18 SCI$_*$ clusters and 18 HM$_*$ clusters.

In Fig. 3 we show normalized likelihood functions for $\sigma_v$
obtained from the SCI$_*$ and HM$_*$ subsamples.  While the
results for the SCI sample are essentially unchanged, we see a
marked reduction in the maximum likelihood value of $\sigma_v$
for the HM$_*$ clusters.  We have determined that the
probability of selecting a subset of 18 clusters at random from
the HM sample that results in a maximum likelihood value of
$\sigma_v$ as small or smaller than that of the HM$_*$ sample is
only $0.002$.  This strongly suggests that the suspect clusters
in the HM sample have systematically higher measured velocities
relative to their reported errors than the other clusters in the
sample.  Given that the excised clusters were chosen due to
ambiguities in their galaxy samples, it seems likely that these
ambiguities have resulted in contributions to their measured
velocities from unaccounted-for errors.  The fact that the SCI
results are not changed significantly by the removal of the
suspect clusters is probably due to the strict membership
criteria employed by Giovanelli et al. (1997) in the selection
of their cluster galaxy samples.  The use of these strict
criteria lowers the probability of incorrect membership
assignment and helps to ensure that each galaxy sample describes
a single collapsed object, thus reducing the possible errors in
the velocities measured for the 6 suspect clusters.

\begin{figure}
\psfig{figure=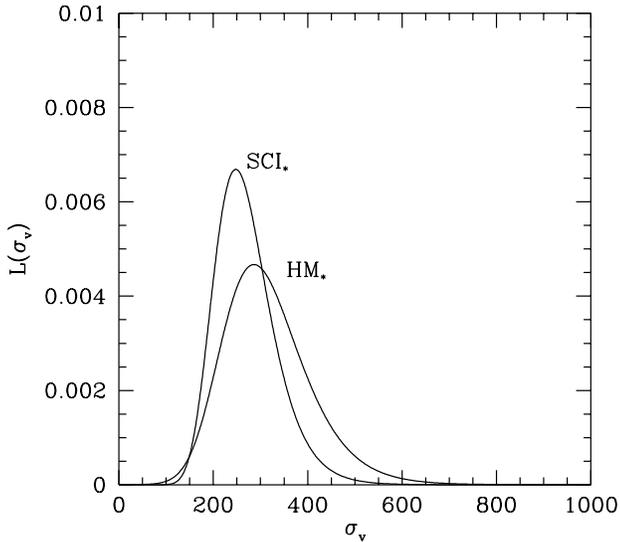,height=3.6in,angle=0}
 \caption{The normalized likelihood functions for $\sigma_v$ obtained from
 the SCI$_*$ and HM$_*$ samples after the removal of suspect clusters. }
\end{figure}

With the suspect clusters removed, the two subsamples are in
good agreement, both favoring a relatively low value of
$\sigma_v$.  The SC$_*$ subsample, which has smaller errors than
the HM$_*$ subsample, yields $\sigma_v= 247^{+123}_{-80}{\rm\
km\ s}^{-1}$ (here and below errors are quoted at the $90\%$
confidence level).  Making the assumptions about the power
spectrum outlined above, this gives $\Omega_0^{0.6}\sigma_8=
0.42^{+0.21}_{-0.14}$, in good agreement with a previous
analysis of the SCI data by Borgani et al. (1997).  We get only
a slight improvement in the accuracy of our result by combining
the SC$_*$ and HM$_*$ samples; keeping the measurement with the
smaller error for clusters present in both results in a combined
sample of 28 clusters yielding $\sigma_v= 257^{+108}_{-76}{\rm\
km\ s}^{-1}$ and $\Omega_0^{0.6}\sigma_8= 0.44^{+0.19}_{-0.13}$.
Using the likelihood function we find that the combined sample
is inconsistent with the value of $489{\rm\ km\ s}^{-1}$ predicted
for standard $\Omega_0=1$ CDM normalized to COBE at the $99.7\%$
confidence level.

\section{Discussion}

We have shown that when suspect clusters are removed, the SCI
and the HM samples show excellent agreement, both favoring
relatively low values for the rms peculiar velocity of clusters;
using a combined sample we find $\sigma_v=
257^{+108}_{-76}{\rm\ km\ s}^{-1}$.  If one
makes some reasonable assumptions about the shape of the power
spectrum, this converts to $\Omega_0^{0.6}\sigma_8=
0.44^{+0.19}_{-0.13}$.  These results are in good agreement with
the results of cluster abundance studies, but are inconsistent
with $\Omega_0=1$ CDM normalized to COBE at the
$99.7\%$ confidence level.  

Our results suggest that peculiar velocities calculated for
galaxy samples that are not representative of a single mass
concentration are susceptible to large unaccounted-for errors.
This is perhaps not too surprising, since velocities are
calculated assuming that all of the sample galaxies have a
common distance.  While this assumption is always violated to
some extent due to the non-zero thickness of a cluster, the
presence of multiple mass concentrations or contamination by
foreground or background galaxies makes it much more likely that
the galaxy sample will by distributed anisotropically about the
cluster center, leading to a bias in the measured velocity.
This problem is especially severe for spiral galaxy samples,
since they tend to inhabit the outer regions of clusters.  The
fact that the central cores of clusters consist primarily of
ellipticals makes it likely that as the fundamental-plane
distance measure continues to improve it will become the
preferred method for measuring the peculiar velocity of
clusters.  

It is unclear what implications, if any, these results have for
the Mark III catalog of galaxy peculiar velocities.  Although
the HM cluster data set was used in the calibration of this
catalog, Willick et al. (1995) recognized that not all of the
galaxy samples were well modeled as belonging to a single
cluster and accounted for this in their calibration procedure.


More cluster peculiar velocities are needed in order to improve
the robustness of our results.  In particular, a larger sample
will allow us to study the dependence on cluster selection and
the validity of our approximation that cluster peculiar
velocities are gaussian distributed.  However, our results
underscore the importance of careful galaxy sample selection
when measuring cluster peculiar velocities.

\section*{Acknowledgments}

The author wishes to thank J. Mohr, H. Feldman, G. Wegner, and
W. Freudling for useful discussions.  This work was supported
in part by NSF grant PHY-9453431 and NASA grant NAGW-4720.

\end{document}